\documentclass[twocolumn,showpacs,preprintnumbers]{revtex4}

\usepackage{mathrsfs}
\usepackage{amssymb}
\usepackage{amsmath}
\usepackage{graphicx}

\begin{document}

\title{Generating correlated (2+1)-photon in an active Raman gain medium}

\author{Chun-Hua Yuan$^{1}$, Cui-Ping Lu$^{1}$, Weiping Zhang$^\dag$$^{1}$, and L. Deng$^{2}$}
\affiliation{$^{1}$State Key Laboratory of Precision Spectroscopy,
Department of Physics, East China Normal University, Shanghai
200062, P. R. China\\
$^{2}$Electron and Optical Physics Division, National Institute of
Standards $\&$ Technology, Gaithersburg, Maryland 20899, USA}

\date{\today }

\begin{abstract}
A scheme of generating controllable (2+1) photons in a
double-$\Lambda$ atomic system based on active-Raman-gain is
presented in this paper. Such (2+1) photons can be a potential
candidate to generate a correlated photon pair as one photon of 2
photons acts as a trigger. Our proposal is an alternative approach
to generate photon pairs where the frequencies are tunable for
subsequent atomic system experiments compared to spontaneous
parametric down-conversion case. Compared to other schemes of
generating correlated photon pairs, our scheme exhibits several
features due to the exploit of the stimulated Raman process and
injection-seeding mechanism.
\end{abstract}

\maketitle


\section{Introduction}

Spontaneous parametric down-conversion (SPDC) is a widely used method for
producing correlated and entangled photon pairs. Due to the spontaneous
emission nature of the process, SPDC-based photon sources usually have
limited applications because of broadband spectrum, low efficiency, short
coherence time and coherence length. Different approaches to generation of
photon pairs have been demonstrated experimentally by a few groups \cite%
{Kuzmich,Wal,Matsukevich104,Balic,Kolchin,Thompson,Du,Pan08}. A few years
ago, Kimble and co-workers working with a magneto-optical trap \cite{Kuzmich}
and Lukin and co-workers working with hot atoms \cite{Wal} have shown the
generation of nonclassical photon pairs. Matsukevich and Kuzmich \cite%
{Matsukevich104} realized the nonclassical photon pairs using\ the two
distinct pencil-shaped components of an atomic ensemble.\ Harris and
co-workers \cite{Balic} demonstrated the generation of counter-propagating
paired photons in a double-$\Lambda $ four wave mixing (FWM) scheme where
the anti-Stokes/FWM field is generated under the electromagnetically induced
transparency (EIT) condition. With a modified scheme Kolchin \emph{et al.}
\cite{Kolchin} showed the generation of narrowband photon pairs using a
standing wave produced by a single intense driving laser. Thompson \emph{et
al.} \cite{Thompson} reported the generation of narrowband pairs of photons
from a laser-cooled atomic ensemble inside an optical cavity. Du \emph{et al.%
} \cite{Du} reported the production of biphotons in a two-level atomic
system. Pan and co-workers \cite{Pan08} demonstrated polarization-entangled
photon pairs using only simple linear optical elements and single photons. A
key element of these experiments \cite%
{Kuzmich,Wal,Matsukevich104,Balic,Kolchin,Thompson,Du} is that one photon is
generated by the spontaneous Raman emission process whereas the second
photon is generated via an EIT-assisted two-wave mixing process.
Alternatively, the process can be viewed as a FWM generation with one
spontaneous emission step in the wave mixing loop.

In this paper, we present a far-detuned, active Raman gain (ARG) scheme \cite%
{Payne,Jiang,Deng,Wang} for the generation of a group of correlated (2+1)
photons. Our goal is to present a different and alternative approach to
generate photon pairs where the frequencies are tunable for subsequent
atomic system experiments compared to SPDC case. Especially, when a perfect
single source is on demand our scheme will have potential applications. At
the first glance the scheme reported here is very similar to those studied
in Refs.~\cite{Wal,Kuzmich,Matsukevich104,Kolchin,Balic,Thompson,Du}.
However, it is operated under a very different principle. Instead of relying
on a spontaneous Raman process to generate the first photon, we inject a
probe photon which leads to stimulated Raman generation. Two features are
evident in this scheme: (1)injection seeding of the probe photon leads to
highly directional generation of probe photons and FWM photons, (2)
stimulated Raman process ensures that the more photons appear in the same
frequency mode and the gain can be easily controllable. The first feature
effectively increases detection efficiency in comparison with a process that
relies on spontaneous emission of the first photon. Due to stimulated
emission the generated probe field has the band width and frequency
characteristics exactly as the inject-seeding field. The second feature can
lead to the generation of photon-number Fock state, which will be explained
in Sec.~\ref{Application}. The precisely controllable directions of photons
due to injection-seeding mechanism in our scheme offer a larger flexibility
for different applications than the undetermined directions of photons in
SPDC case.

\begin{figure}[tbp]
\centerline{\includegraphics[scale=0.5,angle=0]{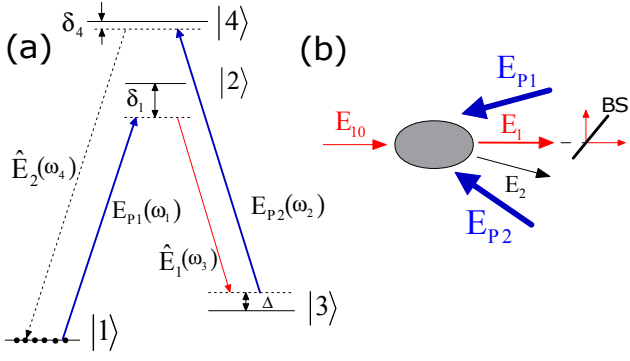}}
\caption{(Color online) (a) Schematic atomic level diagram. $E_{P1}(\protect%
\omega _{1})$ and $E_{P2}(\protect\omega _{2})$ are the pump fields, and $%
\hat{E}_{1}(\protect\omega _{3})$ is the probe field. $\hat{E}_{2}(\protect%
\omega _{4})$ describes the process of another photon $(\protect\omega _{4})$
generated by FWM. (b) Schematic of the experiment for 2+1 photons
generation. }
\label{fig1}
\end{figure}
Carefully choosing the detuning and intensity of the second pump field, our
scheme generates a group of (2+1) photons where two photons are in the same
probe frequency mode whereas one photon is in the FWM frequency mode. The
schematic of the process is shown in Fig.~\ref{fig1}. We consider an
ensemble of identical lifetime broadened four-state atoms initially prepared
in their ground states $|1\rangle $ by suitable optical pumping method. A cw
pump field $(E_{P1})$ couples the ground state $|1\rangle $ to an excited
state $|2\rangle $ with a large one-photon detuning $\delta _{1}$. A second
strong cw laser field $(E_{P2})$ couples the state $|3\rangle $ to an
excited state $|4\rangle $ with the detuning $\Delta +\delta _{4}$. A
single-photon probe field $\hat{E}_{1}$ (with central frequency $\omega _{3}$%
), which couples the state $|2\rangle $ to a lower excited state $|3\rangle $%
, is then introduced to the medium. Consequently, a two-photon Raman
transition can occur, causing an atom to absorb a pump photon and emit a
photon into the probe field frequency mode. The presence of the second pump
field subsequently pumps the atom to state $|4\rangle $ and a FWM process
occurs, leading to the generation of a new photon at the frequency $\omega
_{4}=\omega _{1}-\omega _{3}+\omega _{2}$. In such a process two photons at $%
\omega _{3}$ as well as one photon at $\omega _{4}$ are simultaneously
generated, yielding a group of (2+1) correlated photons \cite{timing}.

The paper is organized as follows. In Sec.~\ref{Model} we first derive the
equations of motion for atomic dynamics and field propagation equations.
With analytical solutions we first examine a limiting case where our scheme
reduces to a well-known single $\Lambda $ gain scheme. We then investigate
group velocities of various propagation modes and corresponding propagation
parameters for the injection-seeded double-$\Lambda $ scheme. In Sec.~\ref%
{pair}, we consider the case where a single-photon quantum probe field is
injected into the medium and we investigate the generation of the two probe
photons and one FWM photon under the condition of weak gain. We also analyze
the two-photon intensity correlation function and coincidence count rate. We
also describe the case where the input probe field is a coherent state and
obtain the single-photon-added coherent state. In Sec.~\ref{Conversion}, we
show the conversion efficiency of our scheme and compare the efficiencies
between our scheme and SPDC case. In Sec.~\ref{Application}, we describe how
to make our scheme be an alternative approach to generate the photon pair
and the possible applications. In Sec.~\ref{Discussion}, we discuss possible
complications due to various ac Stark effects. We also discuss the bandwidth
of the Raman gain and the efficiency and bandwidth of single photon source.
A summary is given in Sec.~\ref{Conclusion}.

\section{Model and analysis}

\label{Model}

\subsection{Theoretical model}

In the electric-dipole and rotating-wave approximations, the interaction
Hamiltonian of $N$ identical four-state atoms interacting with laser fields
(see Fig. 1) is given as
\begin{eqnarray}
&&\hat{H}_{I}^{j}=-\hbar (\Omega _{1}e^{i\vec{k}_{1}\cdot \vec{r}%
_{j}-i\omega _{1}t}\hat{\sigma}_{21}^{j}+\Omega _{2}e^{i\vec{k}_{2}\cdot
\vec{r}_{j}-i\omega _{2}t}\hat{\sigma}_{43}^{j}  \notag  \label{H1} \\
&&+g_{1}\hat{E}_{1}^{(+)}e^{i\vec{k}_{3}\cdot \vec{r}_{j}-i\omega _{3}t}\hat{%
\sigma}_{23}^{j}+g_{2}\hat{E}_{2}^{(+)}e^{i\vec{k}_{4}\cdot \vec{r}%
_{j}-i\omega _{4}t}\hat{\sigma}_{41}^{j}+\mathtt{H.c}),  \notag \\
&&
\end{eqnarray}%
where $\hat{E}_{q}^{(+)}$ is the slowly varying quantum field operator ($%
q=1,2$ are for probe and FWM field, respectively), $\hat{E}_{q}^{(-)}=(\hat{E%
}_{q}^{(+)})^{\dag }$, and $\hat{\sigma}_{kl}^{j}=|k\rangle _{jj}\langle l|$
($k,l=1,2,3,4$). In addition, $2\Omega _{1}=\mu _{21}E_{P1}/\hbar $ and $%
2\Omega _{2}=\mu _{43}E_{P2}/\hbar $ are Rabi frequencies of pump fields
with $\mu _{nn^{\,^{\prime }}}$ being the dipole matrix element for the $%
|n\rangle -|n^{\,^{\prime }}\rangle $ transition and $E_{P1,P2}$ being the
amplitudes of the two classical pump fields. $g_{1}$ and $g_{2}$ are the
atom-photon coupling constants, and $\omega _{m}$ and $\vec{k}_{m}$ $%
(m=1,2,3,4)$ are the carrier frequency and wave vector of the $m-th$ optical
field, respectively.

We define continuum atomic operators $\hat{\sigma}_{\mu\nu}$ by summing over
the individual atoms in a small volume $V$, and introduce slowly varying
atomic operators $\tilde{\sigma}_{\mu\nu}$: $\sigma_{12}=e^{-i\omega_{1}t}%
\tilde{\sigma}_{12},\sigma_{13}=e^{i(\omega_{3}-\omega_{1})t}\tilde{\sigma}%
_{13}=e^{i(\omega_{2}-\omega_{4})t}\tilde{\sigma}_{13},\sigma_{14}=e^{-i%
\omega_{4}t}\tilde{\sigma}_{14}, \sigma_{23}=e^{i\omega_{3}t}\tilde{\sigma}%
_{23},\sigma_{24}=e^{-i(\omega_{4}-\omega_{1})t}\tilde{\sigma}%
_{24}=e^{-i(\omega_{2}-\omega_{3})t}\tilde{\sigma}_{24},\sigma_{34}=e^{-i%
\omega_{2}t}\tilde{\sigma}_{34}$, the equations for the atomic operators in
the Heisenberg picture are
\begin{subequations}
\label{allequations}
\begin{eqnarray}
\dot{\tilde{\sigma}}_{21}&=&-(\gamma_{21}-i\delta_1) \tilde{\sigma}%
_{21}+i\Omega_1^{*}e^{-i\vec{k}_{1}\cdot\vec{r}}(\sigma_{22}-\sigma_{11})
\notag \\
&+&ig_{2}\hat{E}_2^{(-)}e^{-i\vec{k}_{4}\cdot\vec{r}}\tilde{\sigma}%
_{24}-ig_{1}\hat{E}^{(-)}_1 e^{-i\vec{k}_{3}\cdot\vec{r}}\tilde{\sigma}_{31},
\label{equationa} \\
\dot{\tilde{\sigma}}_{34}&=&-(\gamma_{34}+i\delta _{2})\tilde{\sigma}%
_{34}-i\Omega_2e^{i\vec{k}_{2}\cdot\vec {r}}(\sigma_{44}-\sigma_{33})  \notag
\\
&-&ig_{1}\hat{E}_1^{(+)}e^{i\vec{k}_{3}\cdot\vec{r}}\tilde{\sigma}%
_{24}+ig_{2}\hat{E}_2^{(+)} e^{i\vec{k}_{4}\cdot\vec{r}}\tilde{\sigma}_{31},
\label{equationb} \\
\dot{\tilde{\sigma}}_{31}&=&-(\gamma_{31}-i\Delta) \tilde{\sigma}%
_{31}+i\Omega_1^{*}e^{-i\vec{k}_{1}\cdot\vec{r}}\tilde{\sigma}_{32}
-i\Omega_2 e^{i\vec{k}_{2}\cdot\vec{r}}\tilde{\sigma}_{41}  \notag \\
&-&ig_{1}\hat{E}_1^{(+)}\tilde{\sigma}_{21}e^{i\vec {k}_{3}\cdot\vec{r}%
}+ig_{2}\hat{E}_2^{(-)}e^{-i\vec{k}_{4}\cdot\vec{r}}\tilde{\sigma }_{34},
\label{equationc} \\
\dot{\tilde{\sigma}}_{41}&=&-(\gamma_{41}-i\delta _{4})\tilde{\sigma}%
_{41}+i\Omega_1^{*}e^{-i\vec{k}_{1}\cdot\vec{r}}\tilde{\sigma}_{42}
-i\Omega_2^{*}e^{-i\vec{k}_{2}\cdot\vec{r}}\tilde{\sigma} _{31}  \notag \\
&-&ig_{2}\hat{E}_2^{(-)}(\sigma_{11}-\sigma_{44} )e^{-i\vec{k}_{4}\cdot\vec{r%
}},  \label{equationd} \\
\dot{\tilde{\sigma}}_{32}&=&-(\gamma_{32}+i\delta _{3})\tilde{\sigma}%
_{32}+i\Omega_1e^{i\vec {k}_{1}\cdot\vec{r}}\tilde{\sigma}_{31}
-i\Omega_2e^{i\vec{k}_{2}\cdot\vec{r}}\tilde{\sigma}_{42}  \notag \\
&-&ig_{1}\hat{E}_1^{(+)}(\sigma_{22}-\sigma _{33})e^{i\vec{k}_{3}\cdot\vec{r}%
},  \label{equatione} \\
\dot{\tilde{\sigma}}_{42}&=&-[\gamma_{42}-i(\delta_1-\delta_{4})]\tilde{%
\sigma}_{42}+i\Omega_1\tilde{\sigma}_{41}e^{i\vec{k}_{1}\cdot\vec{r}%
}-i\Omega_2^{*}\tilde{\sigma} _{32}  \notag \\
&\times&e^{-i\vec{k}_{2}\cdot\vec {r}}+ig_{1}\hat{E}_1^{(+)}\tilde{\sigma}%
_{43}e^{i\vec {k}_{3}\cdot\vec{r}}  \notag \\
&-&ig_{2}\hat{E}_2^{(-)}\tilde{\sigma }_{12}e^{-i\vec{k}_{4}\cdot\vec{r}},
\label{equationf}
\end{eqnarray}
where $\delta_{1}=\omega_{21}-\omega_{1}, \delta_{2}=\omega_{43}-\omega_{2}$%
, $\delta_{3}=\omega_{23}-\omega_{3}, \delta_{4}=\omega_{41}-\omega_{4}$, $%
\Delta=\delta_3-\delta_1=\delta_2-\delta_4$, and $\omega_{kl}=\omega_{k}-%
\omega_{l}$ is the frequency of the $|k\rangle\leftrightarrow|l\rangle$
transition. $\gamma_{kl}$ ($\gamma_{kl}=\gamma_{lk}$) is the dephasing rate
between state $|k\rangle$ and state $|l\rangle$.

We consider a pencil-shaped atomic ensemble and assume the atomic sample is
optically thin in the transverse direction. Using the slowly varying
envelope and unfocused plane-wave approximations, we obtain the following
propagation equations for the quantum field operators:
\end{subequations}
\begin{equation}
\left[ \left( c\frac{\partial }{\partial z}+\frac{\partial }{\partial t}%
\right) \hat{E}_{1}^{(+)}(z,t)\right] e^{i\vec{k}_{3}\cdot \vec{z}}=ig_{1}N%
\tilde{\sigma}_{32},  \label{propagation1}
\end{equation}%
\begin{equation}
\left[ \left( c\frac{\partial }{\partial z}+\frac{\partial }{\partial t}%
\right) \hat{E}_{2}^{(-)}(z,t)\right] e^{-i\vec{k}_{4}\cdot \vec{z}}=-ig_{2}N%
\tilde{\sigma}_{41},  \label{propagation2}
\end{equation}%
where $N$ is the total number of atoms in the atomic ensemble. Under the
condition that the probe field is much weaker than the pump fields, and
using the assumption that pump fields propagate without depletion, Eqs.~(\ref%
{equationa}) and (\ref{equationf}) can be evaluated adiabatically,
\begin{equation}
\tilde{\sigma}_{21}^{(0)}\approx \frac{\Omega _{1}^{\ast }e^{-i\vec{k}%
_{1}\cdot \vec{z}}}{\delta _{1}+i\gamma _{21}},~\tilde{\sigma}%
_{42}^{(0)}\approx \frac{\Omega _{2}^{\ast }\tilde{\sigma}_{32}e^{-i\vec{k}%
_{2}\cdot \vec{z}}-\Omega _{1}\tilde{\sigma}_{41}e^{i\vec{k}_{1}\cdot \vec{z}%
}}{\delta _{1}+i\gamma _{42}}.  \label{sigma21}
\end{equation}

Considering $\sigma _{22}^{(0)}=\sigma _{33}^{(0)}=\sigma _{44}^{(0)}=\tilde{%
\sigma}_{34}^{(0)}\approx 0$, and substituting these values of the density
matrix elements into Eqs.~(\ref{equationc})-(\ref{equatione}), we obtain the
following set of coupled equations:
\begin{equation}
\left\{ \begin{aligned}
\overset{.}{\tilde{\sigma}}_{31}&=-(\gamma_{31}-i\Delta
)\tilde{\sigma}_{31}+i\Omega_{1}^{\ast}e^{-i\vec{k}_{1}\cdot
\vec{z}}\tilde{\sigma }_{32}\\ &-i\Omega_{2}e^{i\vec{k}_{2}\cdot
\vec{z}}\tilde{\sigma}_{41}-ig_{1}\tilde{\sigma}_{21}%
\hat{E}_{1}^{(+)}(z,t)e^{i\vec{k}_{3}\cdot \vec{z}}, \\
\overset{.}{\tilde{\sigma}}_{41}&=-(\gamma_{41}-i\delta_{4})\tilde{%
\sigma}_{41}+i\Omega_1^{*}e^{-i\vec{k}_{1}\cdot\vec{z}}\tilde{\sigma}_{42}\\
&-i\Omega_{2}^{\ast}e^{-i\vec{k}_{2}\cdot \vec{z}}\tilde{\sigma}_{31}
-ig_{2}\hat{E}_{2}^{(-)}(z,t)e^{-i\vec{k}_{4}\cdot \vec{z}}, \\
\overset{.}{\tilde{\sigma}}_{32}&=-(\gamma_{32}+i\delta_{3})\tilde{%
\sigma}_{32}+i\Omega_{1}\tilde{\sigma}_{31}e^{i\vec{k}_{1}\cdot
\vec{z}}-i\Omega_2e^{i\vec{k}_{2}\cdot\vec{z}}\tilde{\sigma}_{42}.
\end{aligned}\right.  \label{coupled}
\end{equation}

Substituting Eq.~(\ref{sigma21}) into Eq.~(\ref{coupled}), we make the
Fourier transform and obtain the solutions for $\tilde{\sigma}_{32}$ and $%
\tilde{\sigma}_{41}$ in the frequency domain
\begin{eqnarray}
\Sigma _{32} &=&\frac{g_{1}|\Omega _{1}|^{2}}{d_{1}D(\omega )}[(\omega
+d_{4})+\frac{|\Omega _{2}|^{2}-|\Omega _{1}|^{2}}{d_{5}}]\;\hat{\epsilon}%
_{1}(z,\omega )e^{i\vec{k}_{3}\cdot \vec{z}}  \notag \\
&&+\frac{g_{2}\Omega _{1}\Omega _{2}}{D(\omega )d_{5}}(\omega +d_{5})\;\hat{%
\epsilon}_{2}^{\dagger }(z,-\omega )e^{i(\vec{k}_{1}+\vec{k}_{2}-\vec{k}%
_{4})\cdot \vec{z}},  \label{Fourier32} \\
\Sigma _{41} &=&-\frac{g_{1}\Omega _{1}^{\ast }\Omega _{2}^{\ast }}{%
d_{1}D(\omega )}(\omega -d_{3})\;\hat{\epsilon}_{1}(z,\omega )e^{-i(\vec{k}%
_{1}+\vec{k}_{2}-\vec{k}_{3})\cdot \vec{z}}  \notag \\
&&-\frac{g_{2}[(\omega +d_{2})(\omega -d_{3})-|\Omega _{1}|^{2}\ ]}{D(\omega
)}\;\hat{\epsilon}_{2}^{\dagger }(z,-\omega )e^{-i\vec{k}_{4}\cdot \vec{z}},
\notag  \label{Fourier41} \\
&&
\end{eqnarray}%
where $d_{1}=\delta _{1}+i\gamma _{21}$, $d_{2}=\Delta +i\gamma _{31}$, $%
d_{3}=\delta _{3}-i\gamma _{32}$, $d_{4}=\delta _{4}+i\gamma _{41}$, $%
d_{5}=\delta _{1}+i\gamma _{42}$, and $D(\omega )=(\omega
+d_{2})(d_{3}-\omega )(\omega +d_{4})+|\Omega _{1}|^{2}(\omega
+d_{4})+|\Omega _{2}|^{2}(\omega -d_{3})+|\Omega _{1}\Omega _{2}|^{2}(\omega
+2d_{5})/d_{5}^{2}$. Here we ignore some small terms under the condition $%
\delta _{1}(\delta _{3})\gg \delta _{4},\Delta ,\Omega _{1},\Omega _{2}$. In
addition, $\hat{\epsilon}_{j}(z,\omega )=\int_{-\infty }^{\infty }\hat{E}%
_{j}^{(+)}(z,t)e^{i\omega t}dt$, and $\Sigma _{32}(z,\omega )$ and $\Sigma
_{41}(z,\omega )$ are Fourier transforms of $\sigma _{32}(z,t)$ and $\sigma
_{41}(z,t)$, with $\omega $ being the transform variable, respectively.
Making the Fourier transform of Eqs.~(\ref{propagation1}) and (\ref%
{propagation2}) and using Eqs.~(\ref{Fourier32}) and (\ref{Fourier41}) we
obtain
\begin{eqnarray}
\frac{\partial }{\partial z}\hat{\epsilon}_{1}(z,\omega ) &=&\frac{i\omega }{%
c}\hat{\epsilon}_{1}(z,\omega )+iD_{1}(\omega )\hat{\epsilon}_{1}(z,\omega )
\notag \\
&&+iD_{2}(\omega )\hat{\epsilon}_{2}^{\dagger }(z,-\omega )e^{i\Delta \vec{k}%
\cdot \vec{z}},  \label{eq9} \\
\frac{\partial }{\partial z}\hat{\epsilon}_{2}^{\dagger }(z,-\omega ) &=&%
\frac{i\omega }{c}\hat{\epsilon}_{2}^{\dagger }(z,-\omega )+iD_{3}(\omega )%
\hat{\epsilon}_{1}(z,\omega )e^{-i\Delta \vec{k}\cdot \vec{z}}  \notag \\
&&+iD_{4}(\omega )\hat{\epsilon}_{2}^{\dagger }(z,-\omega ),  \label{eq10}
\end{eqnarray}%
where
\begin{eqnarray}
D_{1}(\omega ) &=&\frac{K_{1}|\Omega _{1}|^{2}}{d_{1}d_{5}D(\omega )}%
[(\omega +d_{4})d_{5}+|\Omega _{2}|^{2}-|\Omega _{1}|^{2}],~  \notag \\
D_{2}(\omega ) &=&\frac{K_{12}\Omega _{1}\Omega _{2}(\omega +d_{5})}{%
D(\omega )d_{5}},D_{3}(\omega )=\frac{K_{12}\Omega _{1}^{\ast }\Omega
_{2}^{\ast }(\omega -d_{3})}{D(\omega )d_{1}},  \notag \\
D_{4}(\omega ) &=&\frac{K_{2}[(\omega +d_{2})(\omega -d_{3})-|\Omega
_{1}|^{2}]}{D(\omega )},
\end{eqnarray}%
in which $K_{1}=N|g_{1}|^{2}/c,K_{2}=N|g_{2}|^{2}/c,K_{12}=Ng_{1}g_{2}/c,$
and phase mismatch $\Delta \vec{k}=\vec{k}_{4}-\vec{k}_{3}+\vec{k}_{2}-\vec{k%
}_{1}$. Equations~(\ref{eq9}) and (\ref{eq10}) can be solved analytically
for arbitrary initial conditions. For simplicity, we let the phase mismatch $%
\Delta k=0$. The solutions of Eqs.~(\ref{eq9}) and (\ref{eq10}) are as
follows
\begin{eqnarray}
\hat{\epsilon}_{1}(z,\omega ) &=&\frac{1}{U_{+}-U_{-}}[(U_{+}e^{i\lambda
_{+}z}-U_{-}e^{i\lambda _{-}z})\hat{\epsilon}_{1}(0,\omega )  \notag \\
&+&U_{+}U_{-}(e^{i\lambda _{-}z}-e^{i\lambda _{+}z})\hat{\epsilon}%
_{2}^{\dagger }(0,-\omega )],  \label{a1} \\
\hat{\epsilon}_{2}^{\dagger }(z,-\omega ) &=&\frac{1}{U_{+}-U_{-}}%
[(e^{i\lambda _{+}z}-e^{i\lambda _{-}z})\hat{\epsilon}_{1}(0,\omega )  \notag
\\
&+&(U_{+}e^{i\lambda _{-}z}-U_{-}e^{i\lambda _{+}z})\hat{\epsilon}%
_{2}^{\dagger }(0,-\omega )],  \label{a2}
\end{eqnarray}%
where
\begin{eqnarray}
\lambda _{\pm } &=&\frac{\omega }{c}+\frac{1}{2}[D_{1}(\omega )+D_{4}(\omega
)\mp D_{5}(\omega )],  \notag \\
U_{\pm } &=&\frac{2D_{2}(\omega )}{D_{4}(\omega )-D_{1}(\omega )\mp
D_{5}(\omega )},  \notag \\
D_{5}(\omega ) &=&\sqrt{(D_{1}(\omega )-D_{4}(\omega ))^{2}+4D_{2}(\omega
)D_{3}(\omega )}.
\end{eqnarray}

We focus our attention on the adiabatic regime \cite{Deng1}, where $\lambda
_{\pm }$ and $U_{\pm }$ can be expanded into a rapidly converging power
series of dimensionless transform variable $\zeta =\omega \tau _{p}$ where $%
\tau _{p}$ is the pulse duration of the weak probe field. In this regime
\cite{Deng1}, $U_{\pm }=W_{\pm }+\mathcal{O}(\zeta )$ and $\lambda _{\pm
}=(\lambda _{\pm })_{\zeta =0}+\zeta /\tau _{p}V_{g\pm }+\mathcal{O}(\zeta
^{2})$ can accurately describe the FWM generation and propagation process.
The expansions can be well valid close to the central frequency component of
the field. The inverse Fourier transform of Eqs.~(\ref{a1}) and (\ref{a2})
is given by
\begin{eqnarray}
\hat{E}_{1}^{(+)}(z,t) &=&[A_{1}\hat{E}_{1}^{(+)}(\eta _{+})-A_{2}\hat{E}%
_{2}^{(-)}(\eta _{+})]e^{\beta _{+}z}  \notag  \label{E1} \\
&&+[A_{2}\hat{E}_{2}^{(-)}(\eta _{-})-A_{3}\hat{E}_{1}^{(+)}(\eta
_{-})]e^{\beta _{-}z}, \\
\hat{E}_{2}^{(-)}(z,t) &=&[A\hat{E}_{1}^{(+)}(\eta _{+})-A_{3}\hat{E}%
_{2}^{(-)}(\eta _{+})]e^{\beta _{+}z}  \notag \\
&&+[A_{1}\hat{E}_{2}^{(-)}(\eta _{-})-A\hat{E}_{1}^{(+)}(\eta _{-})]e^{\beta
_{-}z},  \label{E2}
\end{eqnarray}%
where$\ A=1/(W_{+}-W_{-})$, $A_{1}=W_{+}A$, $A_{2}=W_{+}W_{-}A$, $%
A_{3}=W_{-}A$, and $\eta _{\pm }=t-z/V_{g\pm }$, $\beta _{\pm }=i(\lambda
_{\pm })_{\omega =0}$ and
\begin{eqnarray}
W_{\pm } &=&2D_{2}(0)/(D_{4}(0)-D_{1}(0)\mp D_{5}(0)),  \notag \\
\beta _{\pm } &=&i[D_{1}(0)+D_{4}(0)\mp D_{5}(0)]/2,  \notag \\
\frac{1}{{V_{g}}_{\pm }} &=&\frac{1}{c}+\frac{D_{1p}(0)+D_{4p}(0)\mp
D_{5p}(0)}{2},  \label{Vg}
\end{eqnarray}%
in which
\begin{eqnarray}
D_{p}(0) &=&\left\vert \Omega _{1}\right\vert ^{2}+\left\vert \Omega
_{2}\right\vert ^{2}+d_{3}d_{4}-d_{2}d_{4}+d_{2}d_{3},  \notag \\
D_{1p}(0) &=&\frac{K_{1}|\Omega _{1}|^{2}}{d_{1}d_{5}D(0)}%
[d_{5}-(d_{4}d_{5}+|\Omega _{2}|^{2}-|\Omega _{1}|^{2})\frac{D_{p}(0)}{D(0)}%
],  \notag \\
D_{2p}(0) &=&-K_{12}\Omega _{1}\Omega _{2}(D(0)+d_{5}D_{p}(0))/d_{5}D^{2}(0),
\notag \\
D_{3p}(0) &=&K_{12}\Omega_{1}^{\ast }\Omega _{2}^{\ast
}D_{p}(0)(D(0)+d_{3}D_{p}(0))/d_{1}D^{2}(0),  \notag \\
D_{4p}(0) &=&\frac{K_{2}}{D^{2}(0)}[(d_{2}-d_{3})D(0)+D_{p}(0)(d_{2}d_{3}+|%
\Omega _{1}|^{2})],  \notag \\
D_{5p}(0) &=&[(D_{1}(0)-D_{4}(0))(D_{1p}(0)-D_{4p}(0))  \notag \\
&&+2D_{2p}(0)D_{3}(0)+2D_{2}(0)D_{3p}(0)]/\sqrt{D_{5}(0)}.  \notag \\
&&
\end{eqnarray}%
Here $D_{j}(0)\equiv D_{j}(\omega )|_{\omega =0},(j=1,\cdots ,5)$ and $%
D(0)\equiv D(\omega )|_{\omega =0}$. Equations (\ref{E1}) and (\ref{E2})
indicate that in general each frequency component of the probe and the
generated FWM fields contains two propagation modes (wave packets) that
travel with different yet individually matched group velocities. In
addition, both propagation modes (wave packets) retain a pulse shape
identical to that of the input probe field in the adiabatic regime \cite%
{Deng1}.

\subsection{Analysis of group velocities and propagation parameters}

\begin{figure}[tbp]
\centerline{\includegraphics[scale=0.5,angle=0]{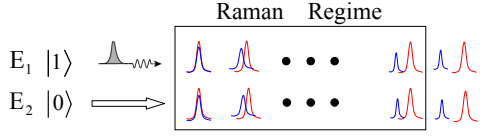}}
\caption{(Color online) The schematic diagram of fields $\hat{E}_{1}$ and $%
\hat{E}_{2}$ propagation. Two propagation modes are characterized by the
different group velocities ${V_{g}}_{\pm }$ where one mode gains and the
other mode attenuates.}
\label{fig2}
\end{figure}
Before a detailed analysis we first consider a limiting case where $\Omega
_{2}=0$ and $g_{2}=0$. In this limit, our scheme reduces to that of a single
$\Lambda $ scheme. Using Eq. (\ref{eq9}) we immediately obtain [note that
when $g_{2}=0$, $D_{2}(\omega )=0$], thus $\hat{\epsilon}_{1}(z,\omega )=%
\hat{\epsilon}_{1}(0,\omega )e^{-i\lambda _{0}z}$ where $\lambda _{0}=\omega
/c+D_{1}(\omega )$. This leads to a single group velocity
\begin{equation}
V_{g_{0}}=\frac{c}{1-\frac{cK_{1}|\Omega _{1}|^{2}}{d_{1}^{2}d_{2}^{2}}}%
\approx \frac{-d_{1}^{2}d_{2}^{2}}{K_{1}|\Omega _{1}|^{2}},\left( \left\vert
\frac{cK_{1}|\Omega _{1}|^{2}}{d_{1}^{2}d_{2}^{2}}\right\vert \gg 1\right) .
\label{e19}
\end{equation}%
Alternatively, one can also obtain this result using Eq. (12). Note that
with $\Omega _{2}=0$ and $g_{2}=0$, we have $D_{2}(\omega )=D_{3}(\omega
)=D_{4}(\omega )=0$, $D_{5}(\omega )=D_{1}(\omega )$, $U_{+}=U_{-}=0$, and $%
\lambda _{-}=\lambda _{0}=\omega /c+D_{1}(\omega )$ and $\hat{\epsilon}%
_{1}(z,\omega )=\hat{\epsilon}_{1}(0,\omega )e^{-i\lambda _{0}z}$. Equation (%
\ref{e19}) is the exact same superluminal group velocity obtained by Payne
and Deng~\cite{Payne} in an atomic amplitude treatment of a single $\Lambda $
active-Raman-gain scheme.

\begin{figure}[tbp]
\centerline{\includegraphics[scale=0.45,angle=0]{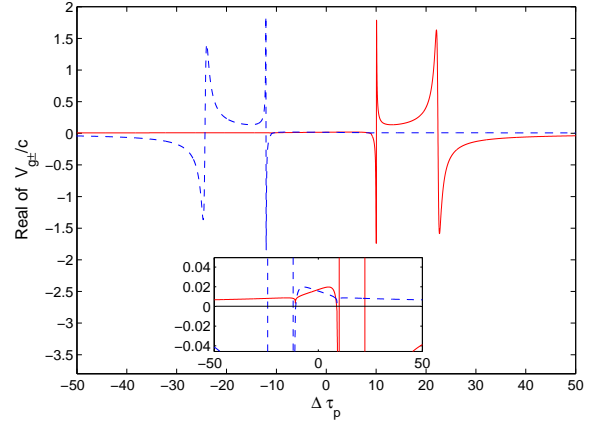}}
\caption{(Color online) Group velocities Re$[V_{g_{+}}/c]$ (solid line) and
Re$[V_{g_{-}}/c]$ (dashed line) as function of dimensionless two-photon
detuning $\Delta \protect\tau_{p}$ for $K_1=K_2=K_{12}=1\times10^{9}$/(m$%
\cdot$s), $|\Omega_1|=\protect\gamma$, $|\Omega_2|=5\protect\gamma$, $%
\protect\tau_{p}=10 ~\protect\mu s$, $\protect\delta_1=100\protect\gamma$, $%
\protect\delta_4=0.1\protect\gamma$, $\protect\gamma _{31}=3\times10^{-5}%
\protect\gamma$, $\protect\gamma_{21}=\protect\gamma _{23}=\protect\gamma%
_{41}= \protect\gamma_{43}=\protect\gamma=10$ MHz, and $\protect\gamma_{42}=2%
\protect\gamma$. The inset shows a magnification of the region for small $%
V_{g_{+}}$ and $V_{g_{-}}$.}
\label{fig3}
\end{figure}

In general, for an injection-seeded double-$\Lambda $ active gain system
presented here, each field consists, as indicated in Eqs.~(\ref{a1}) and~(%
\ref{a2}), of two propagation modes characterized by the eigenvalues $%
\lambda _{\pm }$, and therefore different group velocities $V_{g_{\pm }}$
and decay rates. In the adiabatic regime, the propagation of the fields is
simply governed by Eqs.~(\ref{E1}) and~(\ref{E2}). The schematic diagram of
fields $\hat{E}_{1}$ and $\hat{E}_{2}$ propagation is showed in Fig.~\ref%
{fig2}. For a short propagation distance where the decay of individual mode
is not significant, these modes will overlap each other. Theoretically, for
a sufficiently long propagation distance, however, modes with different
propagation velocities will separate and the fast decay components decay out
completely. Consequently, one obtains a pair of well-matched waves traveling
with the identical group velocity and have the identical decay behavior.

\begin{figure}[tbp]
\centerline{\includegraphics[scale=0.5,angle=0]{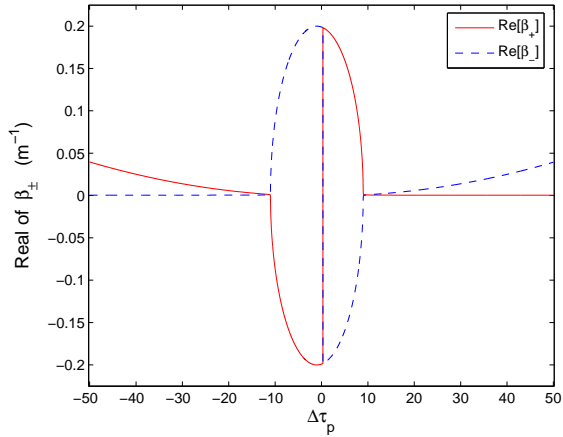}}
\caption{(Color online) The coefficients Re$[\protect\beta_{\pm}]$ versus
the dimensionless two-photon detuning $\Delta \protect\tau_{p}$. The
parameters are chosen as those using in Fig.~\protect\ref{fig3}.}
\label{fig4}
\end{figure}

In Fig.~\ref{fig3} we plot group velocities of each propagation mode (wave
packet) as a function of dimensionless two-photon detuning $\Delta \tau _{p}$
(i.e., the ratio of two-photon detuning to the bandwidth of the probe field)
for a typical cold alkali vapor where states $|2\rangle $ and $|4\rangle $
belong to the same hyperfine manifold. When the two-photon detuning is
large, i.e., $\Delta \tau _{p}>22.4$ or $\Delta \tau _{p}<-24.3$ for this
specific example, one wave-packet mode travels with a negative
(superluminal) group velocity and the other wave-packet mode travels with a
positive (subluminal) group velocity that can be substantially smaller than
the speed of light in vacuum. In this case, the fast and slow modes separate
quickly even before the differential decays become significant.
Consequently, one obtains two group velocity-matched probe-FWM field pairs
arriving at the detector at a delayed time \cite{Deng2}. When the two-photon
detuning is small both propagation modes travel with positive group
velocities. Under the condition where the two group velocities are equal,
one may obtain only one pair of group-velocity matched fields.

To further understand the propagation behavior of probe and generated
fields, we examine the propagation parameters $\beta _{\pm }$ in Eqs. (\ref%
{E1}) and (\ref{E2}). The case of Re[$\beta _{\pm }]>0$ corresponds to gain
whereas Re[$\beta _{\pm }]<0$ indicates field attenuation through
propagation. In Fig.~\ref{fig4} we show these propagation parameters as
functions of the dimensionless two-photon detuning $\Delta \tau _{p}$. In
the region ($-10.8<\Delta \tau _{p}<8.8$) where both propagation modes (wave
packets) travel with positive group velocities one propagation mode
experiences amplification whereas the other propagation mode is attenuated.
Note also that near $\Delta \tau _{p}\approx 0.305$ the signs of propagation
constant change and the amplified (attenuated) mode becomes attenuated
(amplified). These features are very different from the conventional EIT
based FWM schemes where probe attenuation and FWM gain always occur. This is
precisely due to the fact that EIT process is based on the weak absorption
of the probe field and stimulated generation of the FWM field. In the case
of active-Raman-gain medium, the probe field serves as an injection seeding
source and it works in a stimulated emission mode. Consequently,
simultaneous gain to both probe and FWM fields is possible. Indeed, if
parameters are chosen such as Re$[\beta _{+}]>0$ and Re$[\beta _{-}]<0$,
then after a sufficient propagation distance, both the probe and FWM fields
have the same gain feature characterized by Re$[\beta _{+}]>0$. The energy
that supports this bi-field increase comes from the two CW classical fields $%
E_{P1}$ and $E_{P2}$.

\section{Generation of grouped and paired photons}

\label{pair} 
In this section, we consider the case where the quantum state of the
injected probe field corresponds to a single-photon wave-packet state \cite%
{Titulaer,Raymer}:
\begin{equation}
|1\rangle _{\varpi }=\int_{-\infty }^{\infty }d\omega ^{^{\prime
}}P_{p1}(\varpi +\omega ^{^{\prime }})\hat{a}^{\dag }(\omega ^{^{\prime
}})|0\rangle ,
\end{equation}%
where the amplitudes $P_{p1}(\varpi +\omega ^{^{\prime }})$ are normalized
such that $\int_{-\infty }^{\infty }d\omega ^{^{\prime }}|P_{p1}(\varpi
+\omega ^{^{\prime }})|^{2}=1$ and the $\varpi $ is central frequency of
wave packet. Hence the initial state for the system is
\begin{equation}
|\psi _{in}\rangle =|1\rangle _{3}|0\rangle _{4},  \label{initial}
\end{equation}%
where subscripts $3$ and $4$ denote single photon wave packet of central
frequency $\omega _{3}$ and $\omega _{4}$, respectively. In general, the
generated quantum state of the system at time $t$ can be expanded in terms
of boson Fock space as
\begin{equation}
|\psi _{out}\rangle =\sum_{nm}\alpha _{nm}(t)|n\rangle _{3}|m\rangle _{4},
\label{state}
\end{equation}%
where $n$ and $m$ denote the photon numbers of the fields with the central
frequency $\omega _{3}$ and $\omega _{4}$, respectively and $%
\sum_{nm}|\alpha _{nm}(t)|^{2}=1$.

When the generated photon numbers $\{n,m\}$ are small enough in the low gain
case, we can work out the coefficients $\{\alpha _{nm}\}$ in terms of the
moments of the probe and FWM field operators with the forms
\begin{eqnarray}
&&\langle \psi _{out}|F(\hat{E}_{i}^{(-)}(0),\hat{E}_{i}^{(+)}(0),\hat{E}%
_{j}^{(-)}(0),  \notag \\
&&\hat{E}_{j}^{(+)}(0),\cdots )|\psi _{out}\rangle =\langle \psi _{in}|F(%
\hat{E}_{i}^{(-)}(L),  \notag \\
&&\hat{E}_{i}^{(+)}(L),\hat{E}_{j}^{(-)}(L),\hat{E}_{j}^{(+)}(L),\cdots
)|\psi _{in}\rangle ,  \label{relation}
\end{eqnarray}%
where $F(\cdots )$ denotes the combinations of products of the field
operators required to calculate the coefficients $\{\alpha _{nm}\}$. $\hat{E}%
_{i}^{(\pm )}(0)$ and $\hat{E}_{i}^{(\pm )}(L)$ are the quantum field
operators at the entrance $z=0$ and at the output end $z=L$, respectively.

\begin{figure}[tbp]
\centerline{\includegraphics[scale=0.45,angle=0]{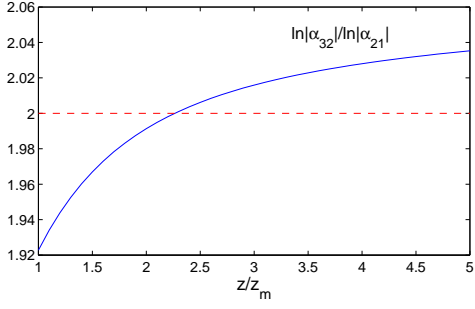}}
\caption{(Color online) The ratio of ln$|\protect\alpha _{32}|/$ln$|\protect%
\alpha _{21}|$ versus the $z/z_{m}$ for $\Delta \protect\tau_{p}=-1$ and $%
z_{m}=1$ cm (solid line). The other parameters are chosen as those using in
Fig.~\protect\ref{fig3}. The dashed line is for the case the ratio is $2$
(SPDC case).}
\label{ratio}
\end{figure}

\subsection{Generation and analysis of (2+1) photon group}

\label{analysis} To generate a (2+1)-group of photons we consider the weak
gain limit in which the injected single-probe photon will generate only one
photon in the probe frequency mode by stimulated Raman process, accompanied
by another photon in the FWM mode. In this case, we only need take account
of the low photon numbers: $n,m\in \{0,1,2\}$ in the expansion of Eq.~(\ref%
{state}). As a result, the final state can be written as
\begin{equation}
|\psi (t)\rangle =\alpha _{10}\left\vert 1\right\rangle _{3}\left\vert
0\right\rangle _{4}+\alpha _{20}\left\vert 2\right\rangle _{3}\left\vert
0\right\rangle _{4}+\alpha _{21}\left\vert 2\right\rangle _{3}\left\vert
1\right\rangle _{4}.  \label{vectorstate}
\end{equation}%
Here, the physical meaning of each term is very clear. $\alpha _{10}$
represents the probability amplitude of the injected single-probe photon not
initiating a stimulated emission. $\alpha _{20}$ describes one photon
generated through the stimulated emission in the probe mode but without
photon generated in the FWM mode. $\alpha _{21}$ is the probability
amplitude of one photon generated through the stimulated emission in the
probe mode, and simultaneously one photon generated in the FWM mode. The
second term in Eq.~(\ref{vectorstate}) only exists for a pump field $\Omega
_{2}$ which is too weak to excite the FWM process. When the second pump
field is strong enough to generate the FWM photon \cite{Notation}, this
amplitude $\alpha _{20}$ tends to zero, and the second term in Eq.~(\ref%
{vectorstate}) vanishes. Further we assume this case to ensure the
generation of a (2+1)-group of photons. Then the state vector simply reduces
to the form
\begin{equation}
|\psi (t)\rangle =|\alpha _{10}||1\rangle _{3}|0\rangle _{4}+e^{i\phi
}|\alpha _{21}||2\rangle _{3}|1\rangle _{4}.  \label{pairphoton}
\end{equation}%
With the help of Eq.~(\ref{relation}), one can work out
\begin{eqnarray}
|\alpha _{10}|^{2} &=&[|A_{1}e^{\beta _{+}L}-A_{3}e^{\beta
_{-}L}|^{2}+|A_{2}|^{2}|e^{\beta _{+}L}-e^{\beta _{-}L}|^{2}  \notag \\
-&4&|A|^{2}|e^{\beta _{+}L}-e^{\beta
_{-}L}|^{2}]P_{p1}^{2}(t-L/V_{g}-(z-L)/c),  \notag \\
|\alpha _{21}|^{2} &=&2|A|^{2}|e^{\beta _{+}L}-e^{\beta
_{-}L}|^{2}P_{p1}^{2}(t-L/V_{g}-(z-L)/c),  \notag \\
\phi &=&\arg [(A_{1}e^{\beta _{+}L}-A_{3}e^{\beta _{-}L})(e^{\beta
_{+}L}-e^{\beta _{-}L})^{\ast }A^{\ast }],  \notag \\
&&
\end{eqnarray}%
where we consider the case with the group velocities $V_{g+}=V_{g-}=V_{g}$,
and the pulse is assumed to maintain its shape with the profile $%
P_{p1}(t-L/V_{g}-(z-L)/c)$ during propagation. It is worth of pointing out
that for a slightly higher gain, the higher order terms, such as $\alpha
_{32}|3\rangle _{3}|2\rangle _{4}$, describing the multiphoton processes
will appear in Eq.~(\ref{pairphoton}). Similarly, multiphoton processes also
exist in the SPDC case in the regime of high gain. For an ideal SPDC case,
the quantum state of photons can be expanded in the form $|\psi \rangle
=|0\rangle +g|1_{i},1_{s}\rangle +g^{2}|2_{i},2_{s}\rangle +\cdots $, and
the probability amplitude of two-photon pairs reduces in the square law for $%
g<<1 $. For comparison, we numerically estimate the corresponding
probability amplitude $|\alpha _{32}|$ in our case. The result shown in Fig.~%
\ref{ratio} indicates that the probability amplitude $|\alpha _{32}|$ has a
faster reduction than the square law for a length $z>2.5z_{m}$. This means
that our scheme using stimulated Raman process with injection-seeding
mechanism is slightly better than the SPDC case for compressing the
multiphoton processes, and hence appropriate for generation of single-photon
pair.

\subsection{Two-photon intensity correlation function and coincidence count
rate}

To show the time correlation properties of the generated photon pairs, we
now work out the Glauber intensity correlation function between the paired
photons $(\omega _{3},~\omega _{4})$ with a time delay $\tau _{d}$,
\begin{equation}
G_{E_{1}-E_{2}}^{(2)}(\tau _{d})= \langle \hat{E}_{1}^{(-)}(t)\hat{E}%
_{2}^{(-)}(t+\tau _{d})\hat{E}_{2}^{(+)}(t+\tau _{d})\hat{E}%
_{1}^{(+)}(t)\rangle.
\end{equation}
Note that the correlation function is calculated for the state of two
photons (one in probe mode and the other in FWM mode), not the state of
three photons. The values $\langle $ $\rangle $ is average of the initial
state $|\psi _{in}\rangle =|1\rangle _{3}|0\rangle _{4}=|0\rangle _{4}\int
d\omega P_{p1}(-\omega )\epsilon _{1}^{\dag }(0,-\omega )|0\rangle _{3}$
where $P_{p1}(\omega )=1/\sqrt{2\pi }\int_{-\infty }^{\infty }dte^{i\omega
t}P_{p1}(t)$ is the pulse shape function of the input photon.

Using the fields operators\ of Eqs.~(\ref{a1}) and (\ref{a2}), one can work
out $G_{E_{1}-E_{2}}^{(2)}(\tau _{d})$. In order to denote the result
simply, Eqs.~(\ref{a1}) and (\ref{a2}) can be written as
\begin{eqnarray}
\hat{\epsilon}_{1}(z,\omega ) &=&R_{1}\hat{\epsilon}_{10}(0,\omega )+S_{1}%
\hat{\epsilon}_{20}^{\dagger }(0,-\omega ), \\
\hat{\epsilon}_{2}^{\dagger }(z,-\omega ) &=&R_{2}\hat{\epsilon}%
_{10}(0,\omega )+S_{2}\hat{\epsilon}_{20}^{\dagger }(0,-\omega ),
\end{eqnarray}%
and the intensity correlation function $G_{E_{1}-E_{2}}^{(2)}(\tau _{d})$
has been derived as
\begin{eqnarray}
&G&_{E_{1}-E_{2}}^{(2)}(\tau _{d})=\int \int \int \int d\omega _{1}d\omega
_{2}d\omega _{3}d\omega _{4}e^{-i\omega _{1}t}  \notag \\
&&\times e^{-i\omega _{2}(t+\tau _{d})}e^{-i\omega _{3}(t+\tau
_{d})}e^{-i\omega _{4}t}\langle \hat{\epsilon}_{1}^{\dagger }(z,-\omega _{1})
\notag \\
&&\times \hat{\epsilon}_{2}^{\dagger }(z,-\omega _{2})\hat{\epsilon}%
_{2}(z,\omega _{3})\hat{\epsilon}_{1}(z,\omega _{4})\rangle  \notag \\
&=&[\int d\omega P_{p1}^{2}(\omega )|R_{1}|^{2}+\int d\omega
|S_{1}|^{2}]\int d\omega |R_{2}|^{2}  \notag \\
&&+\int d\omega P_{p1}^{2}|R_{2}|^{2}\int d\omega |S_{1}|^{2}+|\int d\omega
e^{i\omega \tau _{d}}S_{1}^{\ast }S_{2}|^{2}  \notag \\
&&+\int d\omega e^{-i\omega \tau _{d}}P_{p1}^{2}(\omega )R_{1}^{\ast
}R_{2}\times \int d\omega e^{-i\omega \tau _{d}}S_{1}S_{2}^{\ast }  \notag \\
&&+\int d\omega e^{i\omega \tau _{d}}P_{p1}^{2}(\omega )R_{1}R_{2}^{\ast
}\times \int d\omega e^{i\omega \tau _{d}}S_{1}^{\ast }S_{2}  \notag \\
&\approx &G_{E_{1}}^{(1)}(0)G_{E_{2}}^{(1)}(0)+|\int d\omega e^{i\omega \tau
_{d}}S_{1}^{\ast }S_{2}|^{2}  \notag \\
&&+\int d\omega e^{-i\omega \tau _{d}}P_{p1}^{2}(\omega )R_{1}^{\ast
}R_{2}\times \int d\omega e^{-i\omega \tau _{d}}S_{1}S_{2}^{\ast }  \notag \\
&&+\int d\omega e^{i\omega \tau _{d}}P_{p1}^{2}(\omega )R_{1}R_{2}^{\ast
}\times \int d\omega e^{i\omega \tau _{d}}S_{1}^{\ast }S_{2}.
\end{eqnarray}%
The second-order normalized intensity correlation function is $%
g_{E_{1}-E_{2}}^{(2)}(\tau _{d})=G_{E_{1}-E_{2}}^{(2)}(\tau
_{d})/G_{E_{1}}^{(1)}(0)G_{E_{2}}^{(1)}(0)$ where the peak value of
normalized $g_{E_{1}-E_{2}}^{(2)}\gg 1$, the antibunching nature, which
implies a finite time delay for the emission of the second quantum field
photon $(\omega _{4})$. In other words, a nonclassical photon pair composed
of frequencies $\omega _{3}$ and $\omega _{4}$ is generated in atomic vapors.

\begin{figure}[tbp]
\includegraphics[scale=0.48,angle=0]{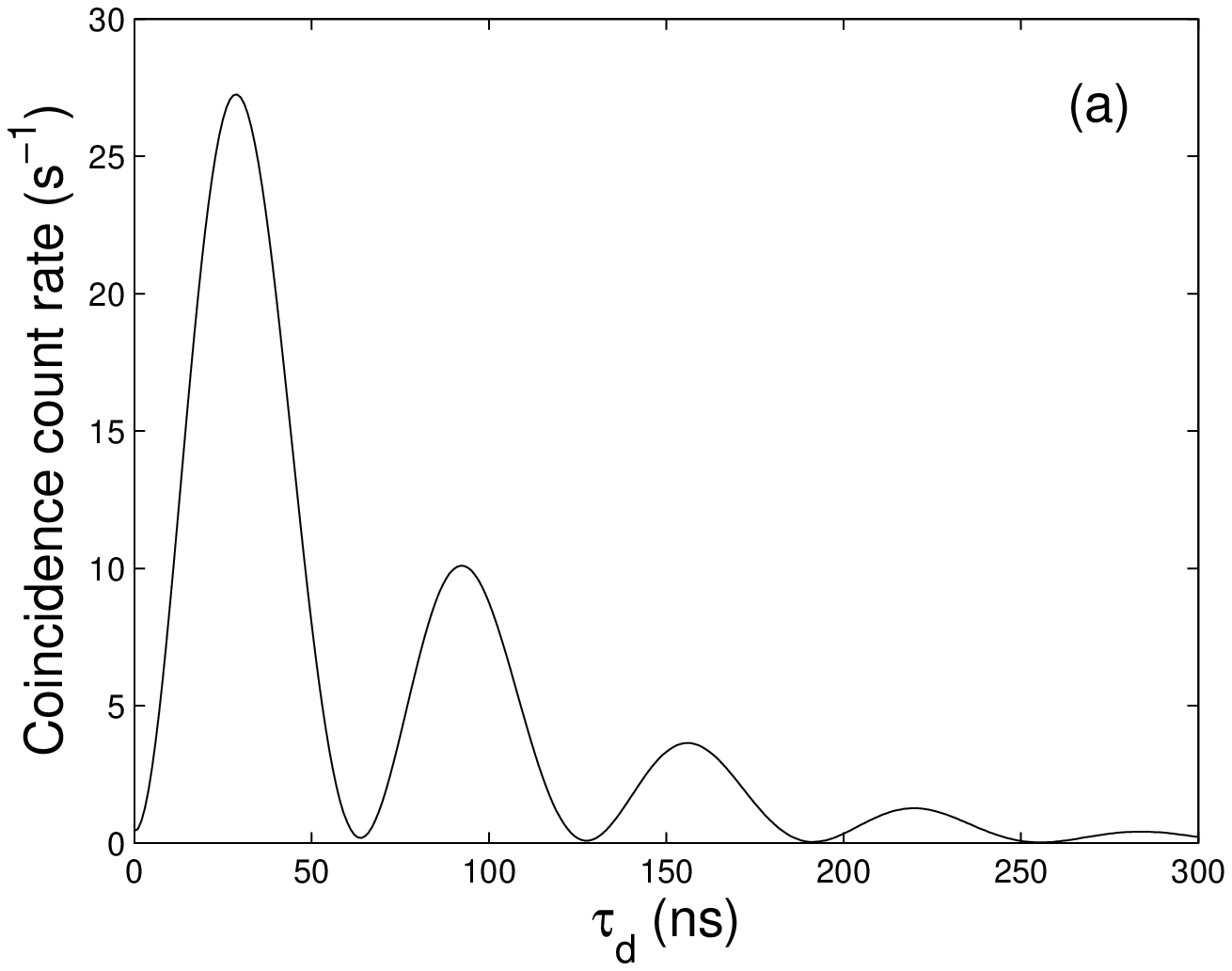} %
\includegraphics[scale=0.48,angle=0]{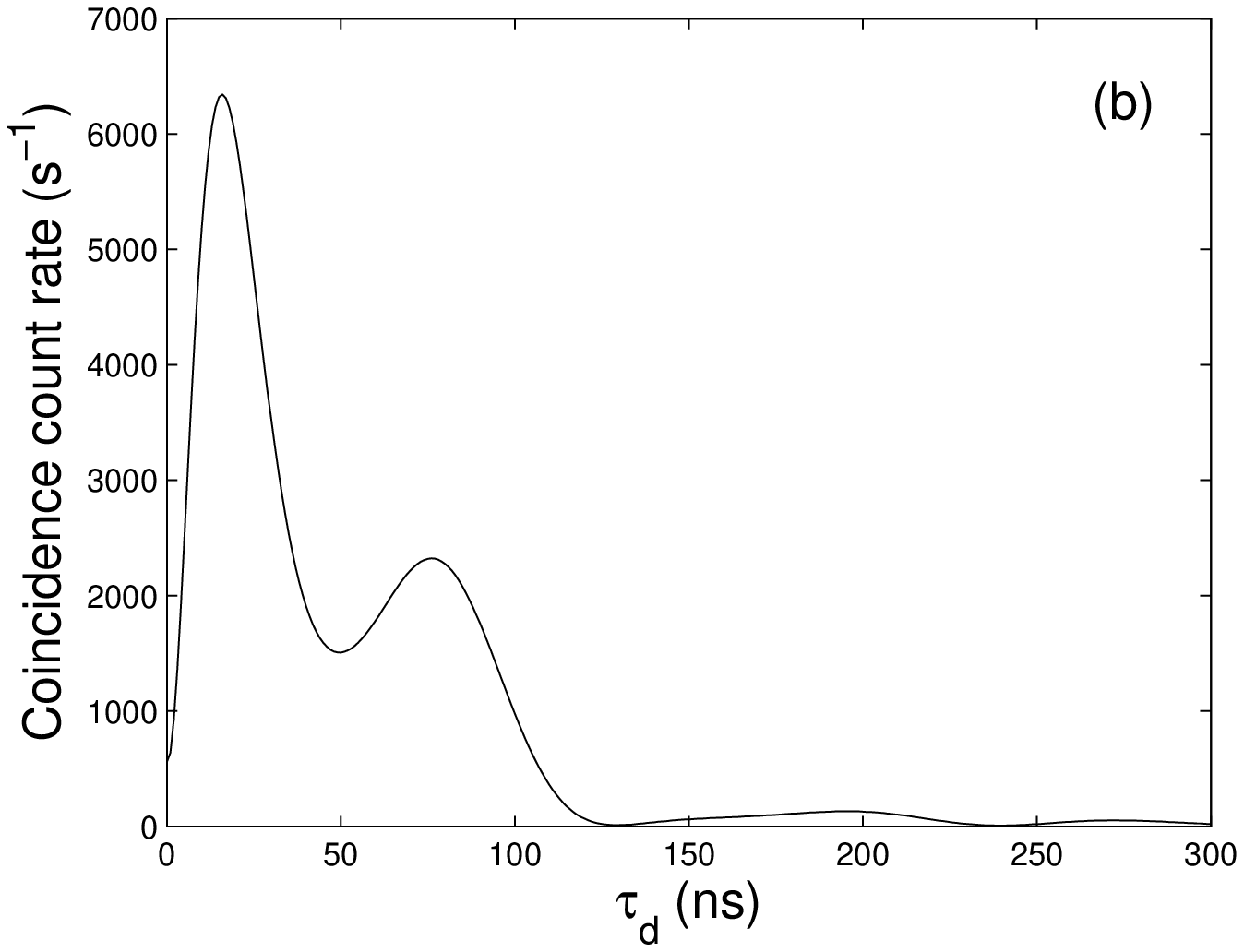}
\caption{Coincidence count rate in a $1$ ns bin versus the delay time $%
\protect\tau _{d}$ for $\Delta \protect\tau _{p}=-1$ and $z/z_{m}=5$ at (a) $%
K=2\times 10^{8}$/(m$\cdot $s) and (b) $K=3\times 10^{9}$/(m$\cdot $s). The
other parameters are chosen as those using in Fig.~\protect\ref{fig3}. }
\label{coincidence}
\end{figure}

The coincidence count rate $R_{c}(\tau _{d})$ is obtained from the intensity
correlation function as $R_{c}(\tau _{d})=\epsilon ^{2}\Delta
TG_{E_{1}-E_{2}}^{(2)}(\tau _{d})$ with a bin size $\Delta T=1$ ns much
smaller than the correlation time. The factor $\epsilon $ accounts for the
photon counter efficiency, the filter transmission, and fiber coupling. In
Fig.~\ref{coincidence}, we show the coincidence count rate $R_{c}(\tau _{d})$
in a $1$ ns bin depending on the $K$. When $K$ is small, which occurs, for
example, at low optical depth, the atomic system behaves like a single atom.
In such a regime the intensity correlation function reveals the damped Rabi
oscillations. By increasing the optical depth of the atomic sample, it
becomes possible to achieve the $E_{1}$-$E_{2}$ correlation function with
shorter time.

\subsection{Single-photon added coherent state}

\begin{figure}[tbp]
\centerline{\includegraphics[scale=0.6,angle=0]{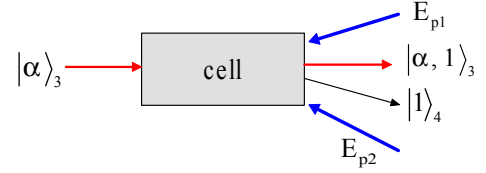}}
\caption{(Color online) The schematic of generating single-photon added
coherent state (SPACS).}
\label{SPACS}
\end{figure}

Another possible extension of the stimulated Raman process with injected
seed is to generate a single-photon added coherent state (SPACS) \cite%
{Agarwal}. For this purpose, we consider injection of a weak coherent state $%
|\alpha \rangle _{3}$ into the medium instead of a probe single photon $%
|1\rangle _{3}$. Here, we also consider the weak gain limit where only one
photon will be generated into the probe field by stimulated Raman process,
accompanied by another photon in the FWM mode. With such an arrangement, we
have the output state
\begin{equation}
\ |\psi \rangle =\beta _{1}|\alpha \rangle _{3}|0\rangle _{4}+\beta
_{2}|\alpha ,1\rangle _{3}|1\rangle _{4}.
\end{equation}%
When only a single photon in the frequency $\omega _{4}$ is detected, the
state $|\psi \rangle $ will collapse to the SPACS $|\alpha ,1\rangle $. The
schematic is shown in Fig.~\ref{SPACS}. Again, our scheme using stimulated
Raman process in atomic ensemble offers a direction-controllable way to
generate a SPACS, compared to the present method demonstrated in the
nonlinear crystal \cite{Zavatta}.

\section{Conversion efficiency}

\label{Conversion} In this section, we come to evaluate the output
conversion efficiency of our photon pair. The intensity of the fields is
given by \cite{Boyd}
\begin{equation}
I=2n(\varepsilon _{0}/\mu _{0})^{1/2}|E|^{2}.
\end{equation}%
where $\varepsilon _{0}=8.85\times 10^{-12}$ F/m, $\mu _{0}=4\pi \times
10^{-7}$ H/m, and $n$ is the refractive index, and $E$ is measured in V/m.
First, we calculate the intensities of pump fields $E_{P1}$ and $E_{P2}$.
The relations of Rabi frequencies of pump fields with the transition matrix
element are $2\Omega _{1}=\mu _{21}E_{P1}/\hbar $ and $2\Omega _{2}=\mu
_{43}E_{P2}/\hbar $, so the intensity of $I_{Pj}$ in the undepleted pump
approximation ($I_{Pj}(z)=I_{Pj}(0)$) is
\begin{equation}
I_{Pj}(0)=8n_{j}(\varepsilon _{0}/\mu _{0})^{1/2}|\hbar \Omega _{j}/\mu
_{j}|^{2},\text{ \ }j=\{1,2\},  \label{pumpintensity1}
\end{equation}%
where $\mu _{1}$ and $\mu _{2}$ denote $\mu _{21}$ and $\mu _{43}$,
respectively.

Next, the intensities of generation fields $E_{1}$ and $E_{2}$ at the
boundary $L$ are
\begin{equation}
I_{E_{j}}=I_{0_{j}}\langle \hat{E}_{j}^{(-)}\hat{E}_{j}^{(+)}\rangle ,\text{
\ \ \ }j=\{1,2\},
\end{equation}%
where $I_{0_{j}}=(\sqrt{\hbar \omega _{2+j}/2\varepsilon _{0}A_{\text{eff}%
}c\tau _{p}})^{2}$ ($j=1,2)$ \cite{Payne03} and $A_{\text{eff}}$\ is the
effective beam cross section,\ and\ the fields $\hat{E}_{1}$ and $\hat{E}%
_{2} $ are described by Eqs.~(\ref{E1}) and (\ref{E2}) and average over
initial the state $|1\rangle _{3}|0\rangle _{4}$. We consider the general
case: only one of two modes $\beta _{+}$\ and $\beta _{-}$ is obtained, such
as $\beta _{+}$, then the peak intensity of the field $I_{E_{1}}$ is
\begin{equation}
I_{E_{1}}(L)=\frac{\hbar \omega _{3}}{2\varepsilon _{0}A_{\text{eff}}c\tau
_{p}}[|A_{1}|^{2}+|A_{2}|^{2}]|e^{\beta _{+}L}|^{2},
\end{equation}%
and the peak intensity of the field $I_{E_{2}}$ under the same condition is
\begin{equation}
I_{E_{2}}(L)\simeq \frac{\hbar \omega _{4}}{\varepsilon _{0}A_{\text{eff}%
}c\tau _{p}}|A|^{2}|e^{\beta _{+}L}|^{2}.
\end{equation}

\begin{figure}[tbp]
\centerline{\includegraphics[scale=1,angle=0]{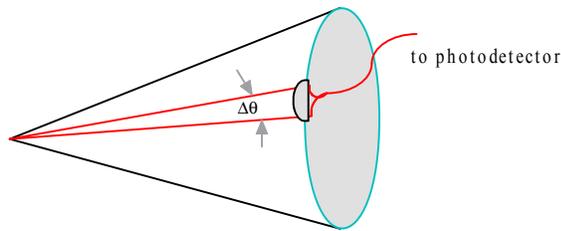}}
\caption{(Color online) The light emitted by SPDC experiment is collected
from a small portion of solid angle.}
\label{cone1}
\end{figure}

The ideal efficiency $\eta _{j}$ ($j=1,2$) for conversion of power from the
pump photons $\omega _{j}$ ($j=1,2$) to the photons $\omega _{2+j}$ ($j=1,2$%
) is
\begin{equation}
\eta _{1}=\frac{P_{3}(L)}{P_{1}(0)}=\frac{I_{E_{1}}(L)}{I_{P1}(0)},\text{ \ }%
\eta _{2}=\frac{P_{4}(L)}{P_{2}(0)}.
\end{equation}%
Considering the efficiency of the present available single-photon source $%
\eta _{s}\approx 9\%$ \cite{Hijlkema}, we can have the total conversion
efficiencies in our scheme $\eta _{\text{tot}_{j}}$ ($j=1,2$) is
\begin{equation}
\eta _{\text{tot}_{j}}=\eta _{j}\eta _{s},\text{ \ }j=\{1,2\}.
\end{equation}

Now, we numerically evaluate the total conversion efficiency $\eta _{\text{%
tot}_{j}}$. As a numeric example with a particular $^{87}$Rb atomic ensemble
using, we assume the hyperfine levels involved: $\{|5^{2}S_{1/2},F=1,m_{F}=1%
\rangle ,~|5^{2}S_{1/2},F=2,m_{F}=1\rangle ,~|5^{2}P_{1/2},F^{\prime
}=2,m_{F^{\prime }}=0\rangle $, ~$|5^{2}P_{3/2},F^{\prime }=2,m_{F^{\prime
}}=2\rangle \}$, which correspond to the levels $\{|1\rangle ,~|3\rangle
,~|2\rangle ,~|4\rangle \}$ in our scheme. As a result, the transition
dipole matrix elements $|\mu _{21}|=2.992$ $ea_{0}/\sqrt{12}$ and $|\mu
_{43}|=4.227$ $ea_{0}/\sqrt{12}$, and the excited transitions $|2\rangle
\rightarrow |3\rangle $ and $|4\rangle \rightarrow |1\rangle $ have the same
Clebsch-Gordon coefficient $1/2$ \cite{Steck}. The spot sizes of the probe
and FWM laser beams $w_{0}$ are assumed to the same with $w_{0}\simeq 10$ $%
\mu $m in our calculation. The dimensionless two-photon detuning\ is chosen
as $z=5z_{m}$ and $\Delta \tau _{p}=3$ corresponding to $\beta _{+}$ mode
and the other parameters are chosen as the same as those using in the Fig. %
\ref{fig3}. Finally, we have the total conversion efficiencies for the
atomic ensemble,
\begin{equation}
\eta _{\text{tot}_{1}}\approx 5.919\times 10^{-8}/\text{cm},  \label{effcon1}
\end{equation}%
and
\begin{equation}
\eta _{\text{tot}_{2}}\approx 4.819\times 10^{-9}/\text{cm}.  \label{effcon2}
\end{equation}%
For comparison, we also examine the conversion efficiency for the
generation of photon pairs in SPDC case. A SPDC process is described
by the Hamiltonian, $\hat{H}=i\hbar \eta A_{p}\hat{a}_{s}^{\dag
}\hat{a}_{i}^{\dag }+$H.c.$,$with a strong classical field $A_{p}$\
and the weak signal (idler) field $\hat{a}_{s}^{\dag }$
($\hat{a}_{i}^{\dag }$). The output state in SPDC case can be
approximated as $|\psi (t)\rangle =|0\rangle +g|1_{s},1_{i}\rangle $
for $g=\eta A_{p}t<<1$. The conversion efficiency of pump photons
into correlated photon pairs integrated over all emission directions
is on the order of $3\times 10^{-8}$ mm$^{-1}$sr$^{-1}$ for a
typical nonlinear material \cite{Klyshko,Ling08}. However, in
practical operation, one photodetector can only collect photons from
a small portion
of solid angle as shown in Fig. \ref{cone1}. For example, in the experiment%
\cite{Ling08}, the collection angle is $3.3\times 10^{-5}$ sr. This leads to
a realistic conversion efficiency is about $1\times 10^{-12}$ mm$^{-1}$ of
crystal length. Taking account of the reduction of efficiency due to
collection angle in real detection for SPDC case, it is evident that the
overall conversion efficiency in our case will be higher due to the
determined direction of emitted photons by stimulated process. In this
sense, our scheme can have some merits over SPDC case and provide a
different method to generate photon pairs using atomic system.
\begin{figure}[tbp]
\centerline{\includegraphics[scale=0.45,angle=0]{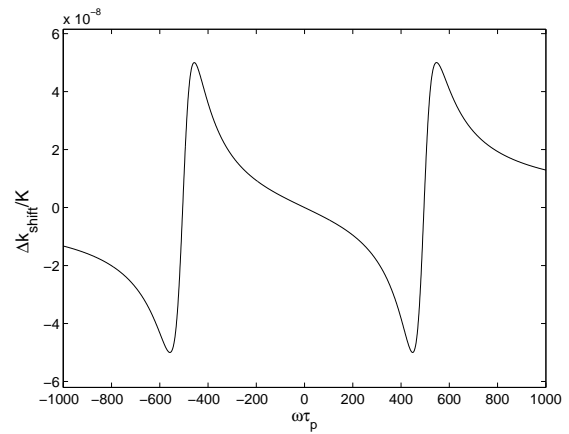}}
\caption{Phase mismatching induced by Stark shift $\Delta k_{\mathtt{shift}}$
as function of detuning $\protect\omega \protect\tau _{p}$ for $\Delta
\protect\tau _{p}=-1$. The other parameters are chosen as those using in
Fig.~\protect\ref{fig3}.}
\label{fig9}
\end{figure}

\section{ Applications of (2+1) photon source}

\label{Application}
In the section, we explore the possibilities for applications of our scheme.
Firstly, the scheme can be used to generate photon pair due to the existence
of the second term on the right side in Eq.~(\ref{pairphoton}). This term
represents the situation of simultaneous existence of two photons in the
probe frequency mode ($\omega _{3}$) and one photon in the FWM frequency
mode ($\omega _{4}$). Detection of a trigger photon ($\omega _{3}$) will
project this state into $|\phi \rangle =\alpha _{10}|0\rangle _{3}|0\rangle
_{4}+e^{-i\phi }\alpha _{21}|1\rangle _{3}|1\rangle _{4}$. In the weak gain
limit considered in the paper this state $|\phi \rangle $ has a large vacuum
component $\alpha _{10}$ and a small admixture of the two photon state $%
\alpha _{21}$. Such a state shares a great similarity to that generated in
SPDC experiments. Despite such a similarity between our scheme and the SPDC
case, our scheme in generating photon pairs has its own reason for
specialty. For example, the frequency of generated photon pairs can be
widely tuned. The conversion efficiency can be achieved effectively higher
than SPDC case due to the injection-seed mechanism and stimulated Raman
process. The injection-seed mechanism ensures a highly directional photon
emission which can increase the collecting efficiency of photons in the
detection, compared to SPDC case. Stimulated Raman process somehow increases
the emission probability. This is because, for an $N$-photon state input, we
have the term $\hat{a}_{k}^{\dag }\hat{a}_{l}^{\dag }|N\rangle _{k}|0\rangle
_{l}=\sqrt{N+1}|N+1\rangle _{k}|1\rangle _{l}$. As a result, the photon
emission probability is $N+1$ times that of the spontaneous emission
described by ($\hat{a}_{k}^{\dag }\hat{a}_{l}^{\dag }|0\rangle _{k}|0\rangle
_{l}=|1\rangle _{k}|1\rangle _{l}$) \cite{Ou08}. Though the enhancement is
low for our case limited by $N=1$, the generation probability is $2$ times
of that from spontaneous emission. Furthermore, the one-photon emission
enhancement due to stimulated emission was observed by Lamas-Linares \emph{%
et al.} \cite{Lamas-Linares}. On the other hand, what is more important is
that if a perfect single photon is on demand, the conversion efficiency from
pump photons into photon pairs will be significantly improved in our scheme.
Hence our scheme can have the potential to take over the SPDC case if the
single-photon-on-demand source is available in the future.

In additional to the generation of photon pairs, our (2+1) photon source can
also be used to generate photon-number Fock state for quantum communication
and other applications. Recently, Ou \cite{Ou08} presented a scheme with a
number of ideas for efficient conversion between photons. In the scheme, one
of the ideas is to use the four-wave mixing as a two-photon annihilator, the
input of which requires a two-photon Fock state. Our (2+1) photon source can
be a candidate for providing such a two-photon Fock state by projecting the
state $|\psi (t)\rangle $ described in Eq.~(\ref{pairphoton}) into the state
$|2\rangle_{3}$ with the detection of the trigger photon $|1\rangle_{4}$.

\begin{figure}[tbp]
\centerline{\includegraphics[scale=0.45,angle=0]{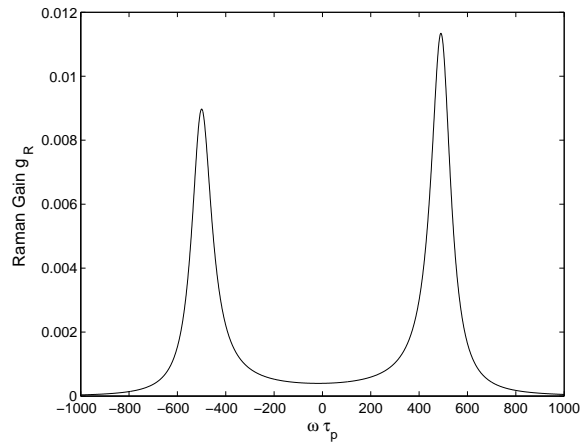}}
\caption{Frequency-dependent gain coefficient for the probe field as
function of detuning $\protect\omega\protect\tau_{p}$ for $\Delta\protect%
\tau _{p}=-1$. The other parameters are chosen as those using in Fig.~%
\protect\ref{fig3}. }
\label{fig10}
\end{figure}

\section{Discussion}

\label{Discussion} In this section we first discuss these ac Stark-type
frequency shifts. From Eqs.~(\ref{eq9}) and~(\ref{eq10}), we obtain the
phase mismatching induced by Stark shift
\begin{equation}
\Delta k_{\mathtt{shift}}=\mathtt{Re}[D_{1}(\omega )-D_{4}(\omega )].
\end{equation}%
Figure~\ref{fig9} shows phase mismatching induced by the Stark shift versus $%
\omega \tau _{p}$ by numerical calculations. It is clear from a comparison
between Figs.~\ref{fig9} and~\ref{fig10} that the mismatching $\Delta k_{%
\mathtt{shift}}$ induced by Stark shift is very small with the exclusion of
two maximal phase mismatches that occur in the regions where the gain
coefficients are maximized. From Fig.~\ref{fig9} we also obtain that the
magnitude of two maximal phase mismatches are small compared with that of
the wave vector of light field, so the angle of departure of field $\hat{E}%
_{2}$ from the direction determine by $\Delta \vec{k}\cdot \vec{z}\approx 0$
is small.

Once the single-photon probe field is introduced, we can generate a
correlated photon pair with probability $|\alpha _{21}|^{2}/2$ when only one
photon $|1\rangle _{3}$ is detected. In order to obtain the desired result,
the Raman gain (by the pump field Rabi frequency $\Omega _{1}$) should be
kept very low to ensure the stimulated generation of one probe photon. Of
course, the pump field Rabi frequency $\Omega _{2}$ should be kept
appropriately to ensure the unit conversion from state $|3\rangle $ to state
$|4\rangle $ which leads to the simultaneous generation of one FWM photon.
Figure~\ref{fig9} depicted the Raman gain in the Fourier transform space for
the parameters used. As expected, the gain is small and rather flat as
required under the condition that the $\omega \tau _{p}$ is small.

We further note that with sufficient $\Omega _{2}$ our scheme has
approximately a EIT-like behavior, as can be seen from Hamiltonian (\ref{H1}%
). In this case, our four-level system is a hybrid system which is composed
of an effective EIT (for the FWM photon generation) and Raman gain (for the
probe photon generation). Specifically, consider the limiting case where $%
\delta _{4}=0$, $\Delta =0$, $|\Omega |=|\Omega _{1}|=|\Omega _{2}|$, and
very large one-photon detuning $\delta _{1}$, we find that the group
velocity $1/{V_{g}}=1/{V_{g}}_{+}=1/{V_{g}}_{-}\approx 1/c+K/(2|\Omega
|^{2}) $ which is EIT-like.

Next, we discuss the bandwidth of the Raman gain and the spatial width of
single photons wave packet. From Fig.~\ref{fig10} we know the magnitude of
Raman gain bandwidth is 10 MHz. The injecting seed single-photon wave-packet
should be microsecond or sub-microsecond pulse. Such single photon source
has so far only been achieved with radiating object (atoms, quantum dots) in
high-finesse microcavities because the efficiency of single photon source is
hard to obtain high in free space due to the light-collecting lens covers
only a fraction of the full 4$\pi $ solid angle. Using a single atom makes
it possible to produce single photons with controlled waveform \cite%
{Kuhn,McKeever,Keller,Hijlkema} and polarization \cite{Wilk}, which allows
realizing deterministic protocols in quantum information science \cite{Wilk2}%
. The deterministic and high efficient single-photon source \cite%
{Hijlkema,Wilk,Matsukevich06,Chen06} can support the present scheme. The
single-photon-generation probability is about $9\%$ by Rempe group \cite%
{Hijlkema}. The low efficiency of single photon sources degrades our scheme.
When a perfect single source is on demand our scheme will have potential
applications.

Finally, we point out that the effect of quantum noises from the atomic
system is ignored since the terms of Langevin-noise operators are excluded
in Eq.~(\ref{allequations}). The ignorance of quantum noises is valid here
as we exploit only the low-gain regime of the active-Raman-gain medium for
the purpose of generation of correlated photons.

\section{Conclusion}

\label{Conclusion}

In conclusion, we have studied a (2+1)-photon generation scheme using a
life-time broadened four-state atomic system. This scheme is based on a
double-$\Lambda $ excitation configuration with the first $\Lambda $ branch
forming an active-Raman-gain medium. This is very different from the
conventional SPDC scheme. Furthermore, it is different from the spontaneous
emission based biphoton generation scheme because of the injection-seeding
mechanism. This injection-seeding technique leads to the highly directional
generation of desired photons, resulting in high detection efficiency. An
important feature of the present scheme is that two identical probe photons,
because of the stimulated Raman emission process due to injection-seeding,
and a FWM photon are generated simultaneously, yielding correlated and
entangled (2+1) photons. Consequently, one of the probe photons can be used
as a coincidence trigger whereas the remaining probe and FWM photons form a
correlated or entangled pair that allows further experimental studies of
paired propagation. In addition, (2+1)-photon source can also be used to
generate photon-number Fock state. Hence if the single-photon-on-demand
source is available in the future, our scheme can have the potential to take
over the SPDC case and have potential applications.

\begin{acknowledgments}
The authors would like to thank Professor Z.Y. Ou and Professor L. You for
helpful discussions. This work was supported by the National Natural Science
Foundation of China under Grants No. 10588402 and No. 10474055, the National
Basic Research Program of China (973 Program) under Grant No. 2006CB921104,
the Science and Technology Commission of Shanghai Municipality under Grants
No. 06JC14026 and No. 05PJ14038, the Program of Shanghai Subject Chief
Scientist under Grant No. 08XD14017, the Program for Changjiang Scholars and
Innovative Research Team in University, Shanghai Leading Academic Discipline
Project under Grant No. B480, the Research Fund for the Doctoral Program of
Higher Education under Grant No. 20040003101. C.H.Y. was supported by the
China Postdoctoral Science Foundation (Grant No. 44021200) and Shanghai
Postdoctoral Scientific Program (Grant No. 44034560). \newline
Email:$^{\dag }$wpzhang@phy.ecnu.edu.cn
\end{acknowledgments}



\begin{thebibliography}{99}
\bibitem{timing} It should be noted that the strong and far-detuned pump
field $E_{P1}$ can still generate a very weak spontaneous Raman emission at
the frequency near the frequency of $\omega_{23}-\delta_1$ when the probe
field is absent (in the vacuum state) in the atomic medium. This process,
however, can be distinguished and discriminated by the usual time-gating
techniques. The Raman gain due to the first pump field should be small in
order to generate only one photon in stimulated process.
\bibitem{Notation} This requires a $\Omega _{2}$ to be on the order of and
larger than $\Omega_{1}$.
\end{thebibliography}
\end{document}